\documentclass[fleqn,preprint,showpacs,preprintnumbers,amsmath,amssymb]{revtex4}
\usepackage{graphicx}
\usepackage{dcolumn}
\usepackage{bm}

\begin{document}

\title{Relativistic degeneracy effect on propagation of arbitrary amplitude ion-acoustic solitons in Thomas-Fermi plasmas}
\author{Abdolrasoul ESFANDYARI-KALEJAHI, Massoud AKBARI-MOGHANJOUGHI and Ehsan SABERIAN}
\affiliation{ Azarbaijan University of
Tarbiat Moallem, Faculty of Sciences,
Department of Physics, 51745-406, Tabriz, Iran}

\date{\today}
\begin{abstract}
Arbitrary amplitude ion-acoustic solitary waves (IASWs) are studied
using Sagdeev-Potential approach in electron-positron-ion plasma
with ultra-relativistic or non-relativistic degenerate electrons and
positrons and the matching criteria of existence of such solitary
waves are numerically investigated. It has been shown that the relativistic
degeneracy of electrons and positrons has significant effects on the
amplitude and the Mach-number range of IASWs. Also it is remarked that
only compressive IASWs can propagate in both non-relativistic and
ultra-relativistic degenerate plasmas.
\\
\\
Keywords: Sagdeev-potential, Ion-acoustic soliton, Degenerate plasma, Relativistic plasma, Fermi statistics
\end{abstract}

\keywords{Sagdeev-potential, Ion-acoustic soliton, Degenerate plasma, Relativistic plasma, Fermi statistics}

\pacs{52.30.Ex, 52.35.-g, 52.35.Fp, 52.35.Mw}
\maketitle

\section{Introduction}

Of the nonlinear excitations, ion-acoustic solitary waves (IASWs)
are of the most important and well-understood characteristics of
plasma environments. Theoretical studies of main properties of these
solitary structures date back to 1961 using Sagdeev
pseud-potentials method \cite{vedenov}. Another method which is
widely used to investigate the collective wave phenomenon in plasma
is the so-called multi-scales perturbation method \cite{salah,
Esfand1, Esfand2, Esfand3, Tiwari1, Tiwari2, Mushtaq}. However, the
latter method, which is based on approximation, is used only for the
small-amplitude treatment of plasma in a state away from
thermodynamic equilibrium. Therefore to obtain a good agreement with
experiments, in this method, one needs to take higher-orders in
perturbation amplitudes. In recent years there have been many
investigations on solitary IASWs as well as solitary electrostatic
waves (ESWs) in diverse plasma environments using Sagdeev
pseudo-potential approach \cite{popel, nejoh, mahmood1, mahmood2,
mahmood3}. The small amplitude propagation and interaction of IASWs with
relativistic degeneracy pressure effects have been recently considered
in Ref. \cite{akbari}.

Among different kinds, pair-plasmas have attracted special attention
in recent years, a special cases of which can be electron-positron
(EP) and electron-positron-ion (EPI) \cite{Shukla, Berezh1, Berezh2,
Berezh3, Rizzato} plasmas. Electron-positron-ion plasma exists in
places such as active galactic nuclei \cite{miller}, pulsar
magnetospheres \cite{michel} and in many dense astronomical
environments, namely, neutron stars and white dwarfs \cite{Ali} and
may play a key role in understanding the beginning and evolution of
our entire universe \cite{rees}. This kind of plasma may also be
practically produced in laboratories \cite{surko1, surko2, greeves,
Berezh}. More specifically when positrons, due to their significant
lifetimes, are used to probe particle transport in Tokamaks, two
component electron-ion (e-i) plasma behaves as three component
(e-p-i) plasma \cite{Surko}. Furthermore, the wave properties such as
stabilities of a two component electron-ion (EI) plasma solitary
excitations may be radically altered by inclusion of low amounts of
positrons.

Owing to their wide applicabilities in micro- and nano-electronic
miniaturization, dense-plasmas is becoming one of the interesting
fields of theoretical as well as experimental fields of plasma
research \cite{Marklund1, Shukla1, Shaikh, Brodin1, Marklund2,
Brodin2, Marklund3}. Dense plasma or the so-called quantum plasma
are characterized by high densities and low temperatures, however, a dense
plasma may be realized in such hot places as planet interiors and white dwarfs \cite{Manfredi}. More
recently, quantum hydrodynamics model has been applied to study the
electron-hole dynamics in semiconductors \cite{Gardner, Markowich}.
The quantum effects in collective behavior of a plasma system
becomes important when the inter-particle distances are comparable
or less than the de Broglie thermal wavelength $\lambda_B = h/(2\pi
m_e k_B T)^{1/2}$ or equivalently when the thermal energies of
plasma species are less than Fermi-energies \cite{Bonitz}. In such
cases the plasma becomes degenerate, in which the plasma ingredients
are under effective influence of Pauli exclusion principle and
classical statistical assumptions break down. Quantum effects also
play important role in the nonlinear processes of white-dwarfs
\cite{Silva}. For instance, for a cold neutron star the densities
can be as high as $10^{15} gm/cm^{-3}$ in the core, which is several
times the density of an atomic nuclei. In extreme conditions such as
the middle of a supernova or the core of a massive white dwarf the
densities can be even catastrophically higher. At these very high
densities the electrons and positrons may become ultra-relativistic
giving rise to the collapse of star under its giant gravitational
force \cite{Chandra, Shapiro}.

The present study is devoted to investigation of propagation of
IASWs in an unmagnetized EPI plasma using Sagdeev pseudo-potential
method in such extreme condition, taking into account the
relativistic degeneracy effects for electrons and positrons. The
basic normalized plasma equations are introduced in section
\ref{equations}. Nonlinear arbitrary-amplitude solution is derived
in section \ref{Sagdeev}. Section \ref{Small amplitude theory}
devotes to short argument about small amplitude IASWs. Numerical
analysis and discussion is given in section \ref{discussion} and
final remarks are presented in section \ref{conclusion}.

\section{Basic plasma state equations}\label{equations}

Consider a dense plasma consisting of electrons, positrons and
positive-ions. Also, suppose that the electrons and positrons follow
the zero-temperature Fermi-gas statistics, while, ions behave as
classical fluid. In such plasma electrons and positrons may be
considered collision-less due to Fermi-blocking process caused by
pauli exclusion principle. Therefore, the semi-classical description
of nonlinear dynamics and interaction of waves in such plasma can be
studied in the framework of conventional hydrodynamics model. The
basic normalized equations describing plasma dynamical state may be
written as
\begin{equation}\label{normal}
\begin{array}{l}
\frac{{\partial {n_i}}}{{\partial t}} + \frac{\partial }{{\partial x}}{n_i}{u_i} = 0, \\
\frac{{\partial {u_i}}}{{\partial t}} + {u_i}\frac{{\partial {u_i}}}{{\partial x}} =  - \frac{{\partial \varphi }}{{\partial x}}, \\
\frac{{{\partial ^2}\varphi }}{{\partial {x^2}}} = {n_e} - {n_p} - {n_i}, \\
\end{array}
\end{equation}
where, electrons and positrons are of Thomas-Fermi type
\begin{equation}\label{thomas}
{n_e} = {(1 + \varphi )^{\frac{3}{2}}},{n_p} = \alpha {(1 - {\sigma _F}\varphi )^{\frac{3}{2}}},\alpha=\frac{n_{p0}}{n_{e0}},\sigma_{F}=\frac{T_{Fe}}{T_{Fp}}. \\
\end{equation}
In obtaining the normalized set of equations following scalings are used
\begin{equation}
x \to \frac{{v_{Fe}}}{{{\omega _{pi}}}}\bar x,t \to \frac{{\bar t}}{{{\omega _{pi}}}},n \to \bar n{n^{(0)}},u \to \bar u{v_{Fe}},\varphi  \to \bar \varphi \frac{{2{k_B}{T_{Fe}}}}{e},
\end{equation}
where, ${\omega _{pi }} = \sqrt {{e^2}n_{e}^{(0)}/{\varepsilon _0}{m_i }}$ and $v_{Fe} = \sqrt {2{k_B}{T_{Fe }}/{m_i }}$ are characteristic plasma-frequency and electron Fermi-speed, respectively, and $n_{e}^{(0)}$ denotes the equilibrium electrons density ($n_e^{(0)} = \frac{{8\pi }}{{3{\hbar ^3}}}p_{Fe}^3$ with $p_{Fe}$ being the electron linear Fermi-momentum).
In a fully degenerate Fermi gas one may write the electron degeneracy pressure in the following general form \cite{Chandra}
\begin{equation}
P = \frac{{\pi m_e^4{c^5}}}{{3{h^3}}}\left[ {r\left( {2{r^2} - 3} \right)\sqrt {1 + {r^2}}  + 3{{\sinh }^{ - 1}}r} \right],
\end{equation}
where, $h$ and $c$ are Plank constant and light-speed, respectively, and $r=p_{Fe}/m_e c$ is the normalized relativity parameter. The electron number density can then be defined in terms of the relativity parameter (${n_e} = \frac{{8\pi m_e^3{c^3}}}{{3{h^3}}}{r^3}$). It is noted that in the limits of very small and very large values of the relativity parameter we obtain
\begin{equation}
P = \left\{ {\begin{array}{*{20}{c}}
{\frac{1}{{20}}{{\left( {\frac{3}{\pi }} \right)}^{\frac{2}{3}}}\frac{{{h^2}}}{{{m_e}}}n_e^{\frac{5}{3}}(r \to 0)}  \\
{\frac{1}{{8}}{{\left( {\frac{3}{\pi }} \right)}^{\frac{1}{3}}}hcn_e^{\frac{4}{3}}(r \to \infty )}  \\
\end{array}} \right\}.
\end{equation}
Therefore, in a three-dimensional \emph{non-relativistic} zero-temperature Fermi-gas for degenerate electrons and positrons from standard definitions we obtain $E_{Fj}=\frac{\hbar^2 k_{Fj}^2}{2m_j}$ ($j=e,p$) or $E_{Fj}\propto n_{j,0}^{2/3}$, which follows that $\sigma_F=\alpha^{-2/3}$. On the other hand, three-dimensional \emph{ultra-relativistic} Fermi-gas, we have $E_{Fj}=c\hbar k_{Fj}$ or $\sigma_F=\alpha^{-1/3}$. It is noted that in our model the inertial ions are always non-relativistic, hence, the non-relativistic hydrodynamics equation has been used in Eqs. (\ref{normal}). Therefore, the Poisson equation reads as
\begin{equation}\label{P}
\frac{{{\partial ^2}\varphi }}{{\partial {x^2}}} = (1 + \varphi)^{\frac{3}{2}}  - \alpha (1 - {\sigma _F}\varphi)^{\frac{3}{2}}  - {n_i}.
\end{equation}
At the equilibrium situation the overall neutrality condition gives rise to the following relation
\begin{equation}
\beta= 1-\alpha,\hspace{3mm} \beta=\frac{n_{i0}}{n_{e0}}.\label{neutral}
\end{equation}

\section{One-dimensional arbitrary-amplitude analysis}\label{Sagdeev}

In this section we derive an appropriate Sagdeev pseudo-potential
describing the dynamics of arbitrary-amplitude IAWs in plasma
containing classical heavy positive ions and inertial-less
relativistic or non-relativistic Thomas-Fermi electrons and
positrons, obeying the three-dimensional distributions in Eqs. (\ref{thomas}).
Using one-dimensional version of Eqs. (\ref{normal}), in a
reduced coordinate $\eta=x-Mt$ (M being Mach number which is a
measure of soliton speed relative to ion-sound speed), from
continuity and momentum equations we obtain
\begin{equation}\label{ni}
{n_i} = \frac{1-\alpha }{{\sqrt {1 - \frac{{2\varphi }}{{{M^2}}}} }},
\end{equation}
where, we have used the fact that $\varphi\rightarrow 0$, $u_i
\rightarrow 0$ and $n_i \rightarrow \beta$ at $\eta\rightarrow
\pm\infty$. Now, substituting Eq. (\ref{ni}) and Eq.
(\ref{thomas}) in Poisson's equation in Eq. (\ref{normal}),
multiplying by $\frac{{d\varphi }}{{d\eta }}$ and integrating with
boundary conditions $\{\varphi,\frac{{d\varphi }}{{d\eta
}}\}\rightarrow 0$ for $\eta\rightarrow \pm\infty$, we derive
\begin{equation}\label{energy}
\frac{1}{2}{\left( {\frac{{d\varphi }}{{d\eta }}} \right)^2} + V(\varphi ) = 0,
\end{equation}
where, the Sagdeev pseudo-potential $V(\varphi)$ reads as
\begin{equation}\label{Sagdeev Eq}
V(\varphi ) = \frac{2}{5}\left[ {1 - {{(1 + \phi )}^{5/2}}} \right]
+ \frac{{2\alpha }}{{5\sigma_F }}\left[ {1 - {{(1 - \sigma_F \phi
)}^{5/2}}} \right] + {M^2}\beta \left[ {1 - \sqrt {1 - \frac{{2\phi
}}{{{M^2}}}} } \right].
\end{equation}
For the reality of $V(\varphi )$ to be ensured we must have
$\varphi\leq M^2/2$ and $\varphi\leq \sigma_F^{-1}$. The possibility of IAWs, therefore, require that
\begin{equation}\label{condition}
{\left. {V(\varphi )} \right|_{\varphi  = 0}} = {\left. {\frac{{dV(\varphi )}}{{d\varphi }}} \right|_{\varphi  = 0}} = 0,{\left. {\frac{{{d^2}V(\varphi )}}{{d{\varphi ^2}}}} \right|_{\varphi  = 0}} < 0,
\end{equation}
it is further required that a $\varphi_m$ exists such that
$V(\varphi_{m})=0$ and for every ${\varphi _m} > \varphi  > 0$ then
$V(\varphi )<0$.

\section{Small amplitude theory}\label{Small amplitude theory}
Let us consider the small-amplitude limit in the above analysis.
Expanding the potential $V(\varphi)$  in (\ref{Sagdeev Eq}) near
$\varphi =0$, we obtain
\begin{equation}
 V(\varphi )=\frac{V''_{0}}{2}\varphi^{2}+\frac{V'''_{0}}{6}\varphi^{3},
\end{equation}
where, $V''_{0}=V''(\varphi=0 )$ and $V'''_{0}=V'''(\varphi=0 )$ are
computed from Eq. (\ref{Sagdeev Eq}) as

\begin{equation}
V''_{0}=-\frac{3}{2}-\frac{3\alpha \sigma_{F}}{2}+\frac{1-\alpha}{M^2},
\end{equation}

\begin{equation}
V'''_{0}=-\frac{3}{4}+\frac{3\alpha \sigma_{F}^{2}}{4\alpha}+\frac{3(1-\alpha)}{M^4}.
\end{equation}

Inserting into Eq(\ref{energy}) and integrating, we obtain (provided
that $V''_{0}<0$ ) a solitary solution in the form (see
\cite{Bertho})

\begin{equation}
\varphi(\eta)=-3\frac{V''_{0}}{V'''_{0}}\frac{1}{\cosh^{2}(\frac{1}{2}\sqrt{-V''_{0}}\eta)}
\end{equation}
This pulse profile is identical to the soliton solution of the
Korteweg-de Vries (KdV) equation, which is obtained by use of the
reductive perturbation method, for example see \cite{Treum}. It is
important to notice that the soliton width $L=2/\sqrt{-V''_{0}}$ and
amplitude $\varphi_{0}= -3V''_{0}/V'''_{0}$ satisfy
$\varphi_{0}L^{2}=12/V'''_{0}=cst.$, as known from the KdV theory.

\section{Numerical analysis and discussion}\label{discussion}

As it was pointed out in section \ref{Sagdeev}, the ion-acoustic
solitary waves (IASWs) exist if the Sagdeev pseudo-potential
satisfies the following conditions: (i) ${\left. {{d^2}V(\varphi
)/d{\varphi ^2}} \right|_{\varphi  = 0}} < 0$, which reveals that
the fixed point $\varphi=0$ is unstable; (ii) $V(\varphi_m)=0$, where,
$\varphi_m$ is the maximum value of $\varphi$; and (iii)
$V(\varphi)<0$ when $\varphi_m>\varphi>0$.

Noticing these conditions, Fig. (1a) and Fig. (1b) show the areas in
$M$-$\alpha$ plane, where: namely, IA solitary waves can exist for
non-relativistic and ultra-relativistic electron-positron
degeneracy, respectively. It is remarked that, the minimum values of
Mach number, $M$, decreases as the fractional positron to electron
number-density ratio, $\alpha$, increases for ultra-relativistic
case (see Fig. 1(b)). Nevertheless, the minimum values of $M$
increases and reaches to a given maximum value then decreases as
$\alpha$ increases (see Fig. 1(a)). On the other hand, the maximum
value of $M$ increases as $\alpha$ increases up to the value
$\alpha\simeq 0.37$ ($\alpha\simeq 0.28$) for  non-relativistic
(ultra-relativistic) case. The maximum value of $M$ decreases as
$\alpha$ increases in the range $1>\alpha>0.37$ ($1>\alpha>0.28$)
for non-relativistic (ultra-relativistic) case.

Comparing Fig. (1a) and Fig. (1b) reveals that both supersonic and
subsonic IASWs can propagate in ultra-relativistic case, whereas,
only subsonic propagations can occur for non-relativistic case.
Although both of supersonic and subsonic IASWs can propagate in
ultra-relativistic degeneracy case, however, the former exist only
for very small range of $\alpha$, for approximately,
$0.07>\alpha>0.4$. For $\alpha=1$ i.e., in the absence of ions,
$V(\varphi)$ does not depend on Mach number, $M$. This case
corresponds to the solution of Poisson equation.

In Fig. (1c) and Fig. (1d), we have numerically
analyzed the Sagdeev potential (Eq. (\ref{Sagdeev})) and
investigated the effects of allowed values of $M$ and $\alpha$ on
the profile of the potential-well for both cases of non-relativistic
and ultra-relativistic degeneracy. It is remarked that for fixed
$\alpha$ ($M$) value, the increase of Much number $M$ ($\alpha$
values) gives rise to an increase of both the potential depth and
amplitude. The energy equation Eq. (9) has been numerically solved
for some values of $\alpha$ and $M$. The potential pulse profiles
has been depicted in Figs. (2a-2c). It is obvious that the potential
profile becomes taller and narrower by increasing $\alpha$ and $M$
for both types of non-relativistic and ultra-relativistic
degeneracy, which is in agreement with above result. Also, we note
that higher pulses are narrower, while shorter are wider, in
agreement with the existing soliton phenomenology. Another important
result is that the amplitude of pulse in non-relativistic case is
higher than ultra-relativistic one for fixed values of $\alpha$ and
$M$ (Fig. 2(c)). Finally, we note that rarefactive IASWs do not exist for both
of non-relativistic and ultra-relativistic degeneracy cases.

\section{Conclusion}\label{conclusion}

The Sagdeev-Potential approach was used to investigate the
propagation of ion-acoustic solitary waves (IASWs) in
electron-positron-ion plasma with ultra-relativistic or
non-relativistic degenerate electrons and positrons. The matching
criteria of existence of such solitary wave were numerically
investigated for both cases of ultra-relativistic or non-relativistic degenerate
electrons and positrons. It is remarked that the characteristics of
nonlinear IASW propagation differ in the mentioned cases. Both
supersonic and subsonic IASWs can propagate in ultra-relativistic
case, whereas, only subsonic propagations can occur for
non-relativistic case. Only compressive IASWs can propagate in both
of desired non-relativistic and ultra-relativistic degenerate
plasmas. Also, it was concluded that, the differences tend to
amplify by moving towards smaller values of fractional
positron-to-electron number-density $\alpha$.
In this work we consider cold ions and the effects of warm ions on
IASWs in such plasmas can be investigated in future.

\newpage

\newpage

\textbf{FIGURE CAPTIONS}

\bigskip

Figure-1

\bigskip

(Color online) The stability regions (dark) of arbitrary amplitude IASWs is shown in $\alpha-M$ plane for non-relativistic (Fig. 1(a)) and ultra-relativistic (Fig. 1(b)) electron/positron degeneracies. Figures 1(c) and 1(d) show the corresponding pseudo-potential dips for varied fractional positron to electron number-densities, $\alpha=0.3$ (blue), $\alpha=0.4$ (grey) and $\alpha=0.5$ (red), and fixed Mach-number in non-relativistic (Fig. 1(c)) and ultra-relativistic (Fig. 1(d)) electron/positron degeneracies, respectively.

\bigskip

Figure-2

\bigskip

(Color online) (a) Ion acoustic solitary wave profiles obtained by numerical solutions of Eq. (\ref{energy}) corresponding to pseudo-potentials shown in Fig. 1(d) which correspond to the values $\alpha=0.25$ (black), $\alpha=0.5$ (blue) and $\alpha=0.75$ (red). (b) Profiles for different supersonic values of Mach-number $M=1$ (black), $M=1.05$ (blue) and $M=1.1$ (red). (c) Comparison between non-relativistic (thin-line in black) and ultra-relativistic (thick-line in blue) electron/positron degeneracies for similar values of $\alpha$ and $M$.

\newpage

\begin{figure}
\resizebox{1\columnwidth}{!}{\includegraphics{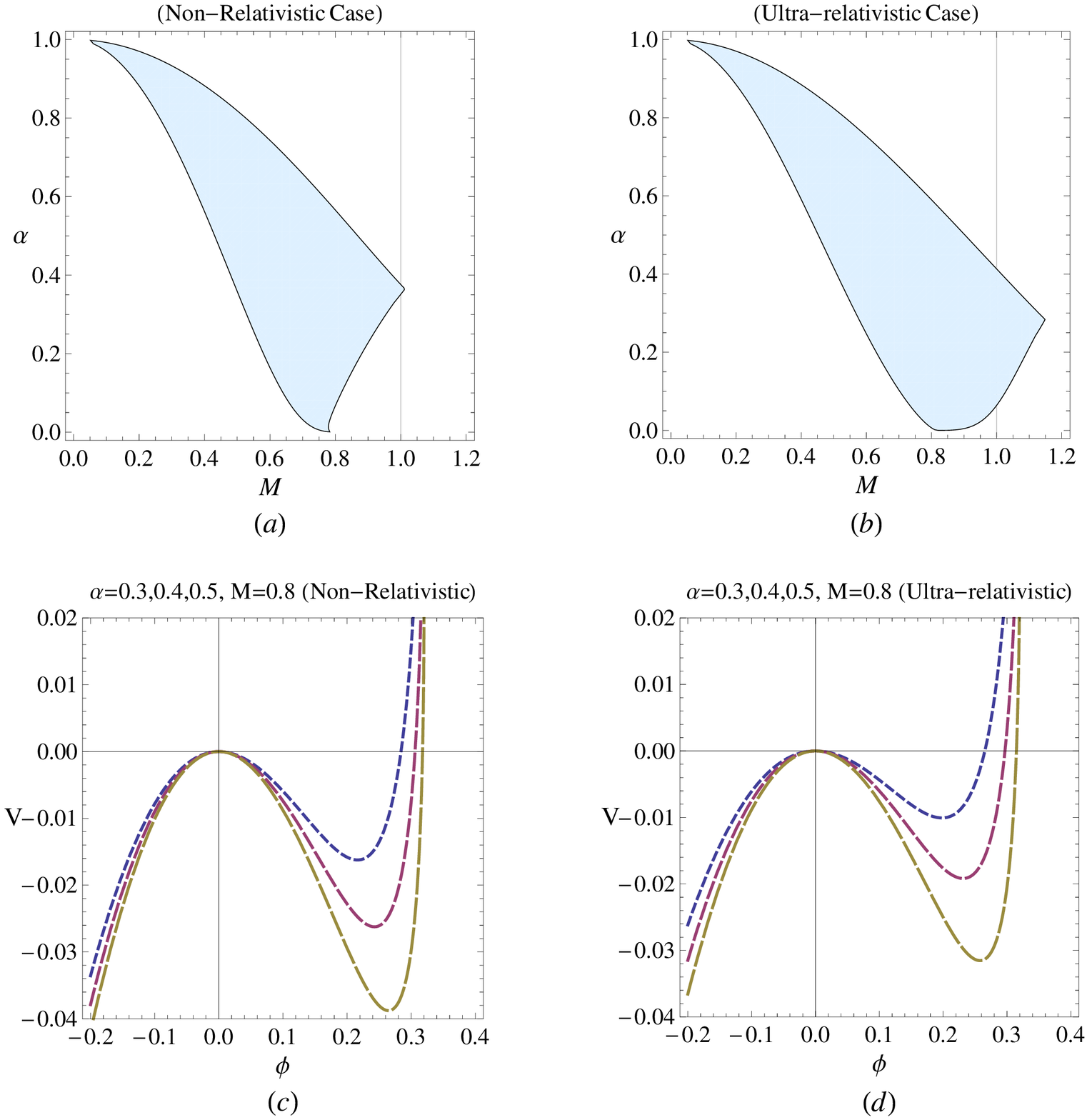}}
\caption{}
\label{fig:1}
\end{figure}

\newpage

\begin{figure}
\resizebox{1\columnwidth}{!}{\includegraphics{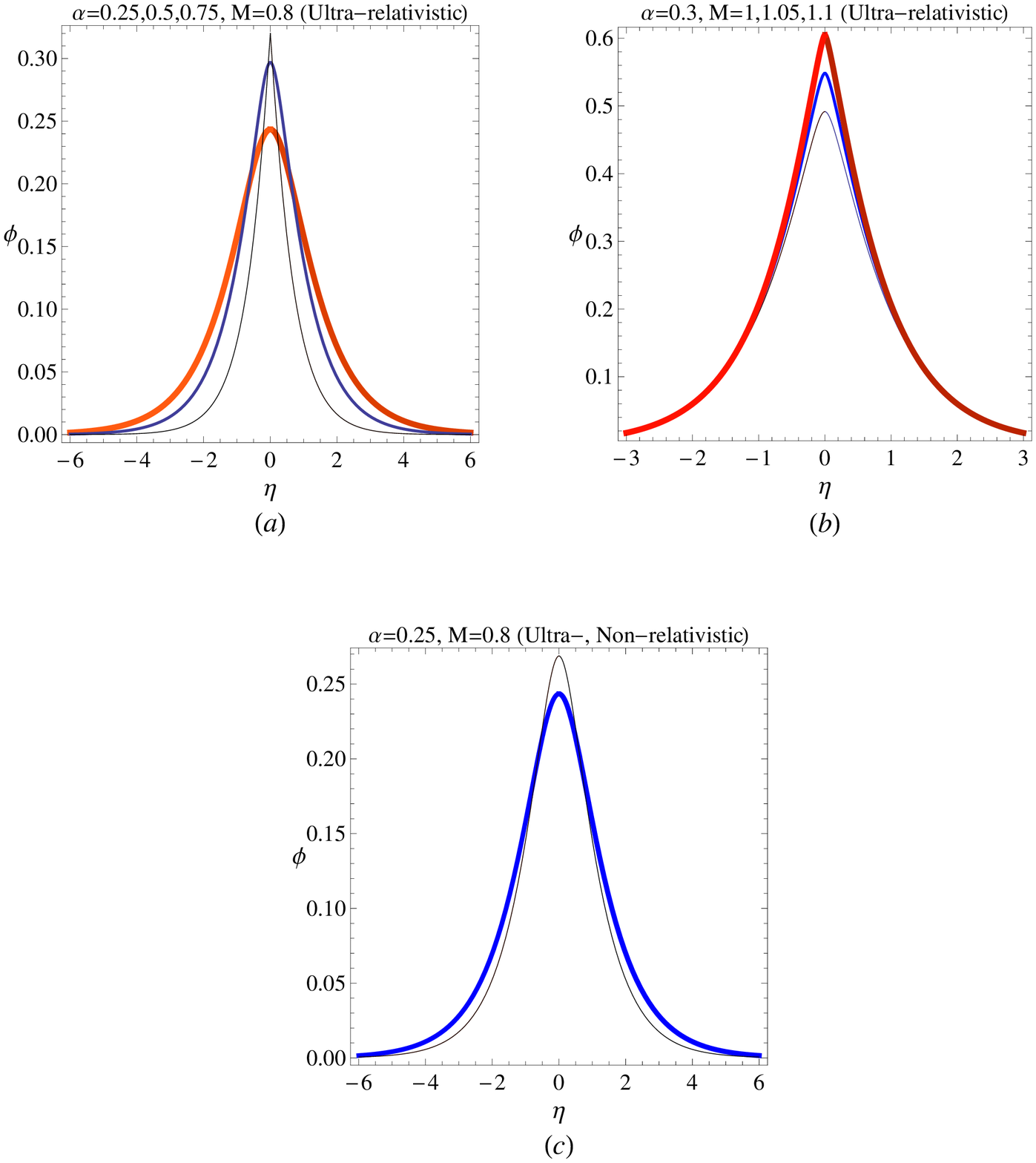}}
\caption{}
\label{fig:2}
\end{figure}

\end{document}